\documentclass[aps,pra,twocolumn,superscriptaddress,showpacs,preprintnumbers,amsmath,amssymb]{revtex4-1}

\usepackage{graphicx}
\usepackage{dcolumn}
\usepackage{bm}
\usepackage{enumerate}
\usepackage{mathrsfs}
\usepackage{amsmath}

\usepackage{epstopdf}
\usepackage[titletoc]{appendix}
\usepackage[colorlinks=true,breaklinks=true,linkcolor=blue,citecolor=blue,urlcolor=blue]{hyperref}
\usepackage{color}
\usepackage{array,mathtools,amssymb,booktabs}
\newcolumntype{C}{>{$}c<{$}}
\AtBeginDocument{
\heavyrulewidth=.08em
\lightrulewidth=.05em
\cmidrulewidth=.03em
\belowrulesep=.65ex
\belowbottomsep=0pt
\aboverulesep=.4ex
\abovetopsep=0pt
\cmidrulesep=\doublerulesep
\cmidrulekern=.5em
\defaultaddspace=.5em
}

\begin{document}

\bibliographystyle{apsrev} 

 \def\rtx@apspra{\class@info{APS journal PRA selected}}

\title{Optical-plug-assisted spin vortex in a $^{87}$Rb dipolar spinor Bose-Einstein condensate}

\author{Hui Tang}
\affiliation{Key Laboratory of Artificial Micro- and Nano-structures of Ministry of Education, School of Physics and Technology, Wuhan University, Wuhan, Hubei 430072, China}
\author{Peng Du}
\affiliation{Key Laboratory of Artificial Micro- and Nano-structures of Ministry of Education, School of Physics and Technology, Wuhan University, Wuhan, Hubei 430072, China}
\author{Lei Jing}
\affiliation{Key Laboratory of Artificial Micro- and Nano-structures of Ministry of Education, School of Physics and Technology, Wuhan University, Wuhan, Hubei 430072, China}
\author{Su Yi}
\affiliation{CAS Key Laboratory of Theoretical Physics, Institute of Theoretical Physics, Chinese Academy of Sciences, P.O. Box 2735, Beijing 100190, China}
\affiliation{School of Physical Sciences \& CAS Center for Excellence in Topological Quantum Computation, University of Chinese Academy of Sciences, Beijing 100049, China}

\author{Wenxian Zhang}
\email[Corresponding email: ]{wxzhang@whu.edu.cn}
\affiliation{Key Laboratory of Artificial Micro- and Nano-structures of Ministry of Education, School of Physics and Technology, Wuhan University, Wuhan, Hubei 430072, China}
\affiliation{Wuhan Institute of Quantum Technology, Wuhan, Hubei 430206, China}

\date{\today}

\begin{abstract}
Generating a spin vortex in a $^{87}$Rb dipolar spinor Bose-Einstein condensate in a controllable way is still experimentally challenging. We propose an experimentally easy and tunable way to produce spin vortex by varying the potential barrier height and the width of an additionally applied optical plug. A topological phase transition occurs  from the trivial single mode approximation phase to the optical-plug-assisted-vortex one, as the barrier height increases and the width lies in an appropriate range. The optical plug causes radial density variation thus the spin vortex is favored by significantly lowering the intrinsic magnetic dipolar energy. A type of coreless spin vortex, different from the conventional polar core vortex,  is predicted by our numerical results. Our proposal removes a major obstacle to investigate the topological phase transition in a $^{87}$Rb dipolar spinor BEC.
\end{abstract}

\maketitle

\section{ Introduction}

Phase transitions are ubiquitous in the classical world~\cite{peters2012elements}. While in the quantum world,  phase transitions are an important and developing branch of quantum theory, especially for the unconventional topological phase transitions and dynamical quantum phase transitions. Although various types of quantum phase transitions have been observed for decades in many quantum systems, such as superconducting phase transitions, magnetic phase transitions, quantum Hall phase transitions~\cite{kamerlingh1911resistance1, gantmakher2010superconductor, majlis2007quantum, PhysRevLett.45.494, sachdev1999quantum}, the quantum phase transitions in a controllable way becomes possible only recently, with the aid of fine-tuning quantum control techniques developed in the last twenty years~\cite{chen2007controllable, luo2017deterministic, PhysRevLett.123.073001}.

The Bose-Einstein condensate (BEC) in alkali atomic gases offers an excellent platform for the experimental investigation of quantum phase transitions, due to its extremely clean environment, macroscopical quantum behaviors, and great controllability. Particularly, when the optical potential was first utilized to trap the atomic gases, a vast unexplored area in the ultracold atomic physics was opened up by releasing the spin degree of freedom~\cite{PhysRevLett.80.2027}. Beautiful theories and extensive experimental investigations of spinor BEC have been developed and carried out latter~\cite{doi:10.1143/JPSJ.67.1822, PhysRevLett.81.742, PhysRevLett.81.5257, PhysRevA.66.011601, RevModPhys.85.1191, KAWAGUCHI2012253, PhysRevA.67.033403, PhysRevLett.87.010404}.

It has been well known that there exist two kinds of spin interaction in a dipolar spinor condensate, the short range and isotropic spin exchange interaction and the long range and anisotropic magnetic dipole-dipole interaction~(MDDI)~\cite{PhysRevLett.93.040403, PhysRevA.66.013607, PhysRevLett.88.170406, PhysRevA.61.051601, lahaye2009physics, PhysRevLett.95.150406, PhysRevLett.100.170403, PhysRevLett.107.190401}. The competition between these two spin interactions drives the system into a rich spin phase diagram, e.g., from the ferromagnetic phase to the antiferromagnetic one by changing the spin exchange interaction, from the spin-uniform  single mode approximation (SMA) phase to the polar-core-vortex (PCV) one by increasing the MDDI strength, and so on~\cite{PhysRevLett.97.020401,PhysRevLett.97.130404, PhysRevA.81.063623}.

Among these phase transitions, the topological phase transition is of particular interest due to its sudden change of the global property at the critical point. It also offers a new paradigm of quantum phase transition. For a metastable state, topological vortices and spin domains were pursued experimentally and numerically with large scale supercomputers through quench dynamics in a $^{87}$Rb dipolar spinor condensate~\cite{sadler2006spontaneous, PhysRevLett.100.170403, PhysRevA.75.013621}. For the ground state, however, the topological phase transition from the SMA to the PCV was only predicted theoretically and confirmed numerically in a dipolar spinor BEC. It has yet been realized in experiments~\cite{PhysRevLett.97.020401,PhysRevLett.97.130404}. The conditions to generate such a ground-state spin vortex are rather challenging, either with highly anisotropic trap potential or with extremely large number of atoms. A more controllable manner and less stringent experimental conditions to realize the topological phase transitions are still seriously in need.

In this paper, we apply an additional optical plug to a $^{87}$Rb dipolar spinor BEC in a highly pancake optical trap~\cite{PhysRevLett.75.3969, PhysRevLett.104.160401, fu2003effect}. By varying the potential barrier height and the width of the optical plug, we expect to realize the topological phase transition in the $^{87}$Rb condensate with less number of atoms and experimentally available trap aspect ratio. With analytical arguments and numerical calculations, we illustrate the phase transition from the SMA to the optical-plug-assisted spin vortex, due to the competition between the MDDI interaction and the kinetic energy with the optical plug. The generated spin vortex is further divided into two cases: a PCV and a flux-closure coreless spin vortex (FCLSV).
Actually, the FCLSV has a similar spin structure to the PCV but the polar core was strongly suppressed by the optical plug. Our results point out a viable and tunable way to experimentally investigate the topological quantum phase transition in a $^{87}$Rb dipolar spinor condensate.

The paper is organized as follows. We describe the nonlocal model with the coupled Gross-Pitaevskii equations (GPEs) for the $^{87}$Rb dipolar spinor BEC in Sec.~\ref{sec:nonlocal}. We sketch the idea and the analytical arguments of generating a spin vortex with the application of an additional optical plug in Sec.~\ref{sec:opv}. Then in Secs.~\ref{sec:trunc} and~\ref{sec:num}, we respectively describe the numerical truncation technique for the accurate calculation of the MDDI in the condensate and present the numerical results which confirm the topological phase transition from the SMA to the spin vortex. Conclusions are given  in Sec.~\ref{sec:conc}.


\section{Nonlocal dipolar model}
\label{sec:nonlocal}

Due to its long range nature of MDDI, a $^{87}$Rb dipolar spin-1 BEC in an optical trap is described by a nonlocal model. The Hamiltonian in the second quantized form is~\cite{PhysRevA.82.043627, PhysRevLett.97.130404, PhysRevA.73.023602, lahaye2009physics},
\begin{equation}
\begin{aligned}
\hat{H}=&\int d\mathbf{r} \left[\hat{\psi}^{\dag}_{\alpha}(\mathbf{r})\left(-\frac{\hbar^2\nabla^2}{2M}+V(\mathbf{r})\right)\hat{\psi}_{\alpha}(\mathbf{r})\right.\\
&+\frac{c_0}{2}\hat{\psi}^{\dag}_{\alpha}(\mathbf{r})\hat{\psi}^{\dag}_{\beta}(\mathbf{r})\hat{\psi}_{\beta}(\mathbf{r})\hat{\psi}_{\alpha}(\mathbf{r})\\
&+\left.\frac{c_2}{2}\hat{\psi}^{\dag}_{\alpha}(\mathbf{r})\hat{\psi}^{\dag}_{\alpha^{\prime}}(\mathbf{r})\mathbf{\textit{\textbf{F}}}_{\alpha\beta}\cdot\mathbf{\textit{\textbf{F}}}_{\alpha^{\prime}\beta^{\prime}}\hat{\psi}_{\beta}(\mathbf{r})\hat{\psi}_{\beta^{\prime}}(\mathbf{r})\right] \\
&+\hat{H}_{dd}
\end{aligned}
\end{equation}
where $\hat{\psi}_{\alpha}$ $(\alpha=0, \pm1)$ represents the annihilation operator in the
magnetic level $\alpha$, $M$ the mass of a $^{87}$Rb atom, and $\mathbf{\textit{\textbf{F}}_{\alpha\beta}}$ the spin-1 matrix. The spin-independent interaction strength is $c_0=4\pi\hbar^{2}(a_0+2a_2)/(3M)$ and spin-dependent exchange interaction is $c_2=4\pi\hbar^{2}(a_2-a_0)/(3M)$, which originate from the two-body $s$-wave scattering with $a_s (s=0,2)$ being the characteristic scattering length of the total spin $s$.

The $V(\mathbf{r})$ is an external harmonic trap with an additional optical plug,
\begin{equation}
\setlength\abovedisplayskip{3pt}
\setlength\belowdisplayskip{3pt}
V(\mathbf{r}) = \frac{1}{2}M\omega_0^{2}(x^2+y^2+\lambda^2z^2) +U_{0}\;e^{-(x^2+y^2)/\sigma^2}
\end{equation}
where $\omega_{0}$ is the harmonic trap angular frequency in the $x$-$y$ plane, $\lambda$  trap aspect ratio, and $U_{0}$ and $\sigma$ are two adjustable parameters characterizing respectively the barrier height and the width of the Gaussian optical plug~\cite{PhysRevLett.75.3969, PhysRevLett.104.160401, fu2003effect}.

The Hamiltonian of the MDDI $\hat{H}_{dd}$ is
\begin{equation}
\setlength\abovedisplayskip{3pt}
\setlength\belowdisplayskip{3pt}
\begin{split}
\hat{H}_{dd}=&\frac{c_{dd}}{2}\int d\mathbf{r}\int d\mathbf{r^{\prime}}\frac{1}{|\mathbf{r-r^{\prime}}|^{3}}\\
\times &[\hat{\psi}_{\alpha}^\dag(\mathbf{r})\hat{\psi}_{\alpha^{\prime}}^\dag(\mathbf{r})\mathbf{\textit{\textbf{F}}}_{\alpha\beta}\cdot\mathbf{\textit{\textbf{F}}_{\alpha^{\prime}\beta^{\prime}}}\hat{\psi}_{\beta}(\mathbf{r})\hat{\psi}_{\beta^{\prime}}(\mathbf{r})\\
&-3\hat{\psi}_{\alpha}^\dag(\mathbf{r})\hat{\psi}_{\alpha^{\prime}}^\dag(\mathbf{r})\left(\mathbf{\textit{\textbf{F}}}_{\alpha\beta}\cdot\mathbf{e}\right)\left(\mathbf{\textit{\textbf{F}}}_{\alpha^{\prime}\beta^{\prime}}\cdot\mathbf{e}\right)\hat{\psi}_{\beta}(\mathbf{r})\hat{\psi}_{\beta^{\prime}}(\mathbf{r})]
\end{split}
\end{equation}
where $c_{dd}=\mu_{0}g_{F}^{2}\mu_{B}^{2}/(4\pi)$ with $\mu_0$ the magnetic permeability of the vacuum, $g_F$ the Land\'e $g$ factor for the $^{87}$Rb atom, and $\mu_B$ the Bohr magneton, and $\mathbf{e}=(\mathbf{r}-\mathbf{r^{\prime}})/|\mathbf{r}-\mathbf{r^{\prime}}|$ is the unit vector along $\mathbf{r}-\mathbf{r^{\prime}}$.

By adopting the standard mean-field approximation~\cite{KAWAGUCHI2012253}, the order parameter (wave function) of the dipolar spin$-1$ BEC becomes $\Psi(\textbf{r})=(\psi_{1}(\textbf{r}),\psi_{0}(\textbf{r}),\psi_{-1}(\textbf{r}))^{T}$.
The dynamics of the system is described by the following nonlocal GPEs~\cite{PhysRevA.93.053602, PhysRevA.102.013305},
\begin{equation}\label{eq:gpes}
\setlength\abovedisplayskip{3pt}
\setlength\belowdisplayskip{3pt}
i\hbar\frac{\partial\psi_{\alpha}(\mathbf{r})}{\partial t}= (T+V+c_{0}n)\psi_{\alpha}+\textbf{B}_{e} \cdot \textbf{F}_{\alpha\beta}\psi_{\beta}
\end{equation}
where $T=-\hbar^2\nabla^2/(2M)$ and $n=\sum_{\alpha}|\psi_{\alpha}|^2$ the total density. All spin-dependent interactions are considered as an effective magnetic field,
\begin{equation}\label{eq:eff}
\textbf{B}_{e}({\textbf{r}})=c_{2}\textit{\textbf{f}}(\textbf{r})+c_{dd}\textbf{D}(\textbf{r})
\end{equation}
with $\textit{\textbf{f}}(\textbf{r})=\sum_{\alpha\beta}\psi_{\alpha}^\ast(\textit{\textbf{F}}_{\alpha\beta})\psi_{\beta}
$ is the spin density and $\textbf{D}(\textbf{r})$ denotes dipolar contribution with its $\mu (\mu=x,y,z)$ component defined as,
\begin{equation}
\setlength\abovedisplayskip{3pt}
\setlength\belowdisplayskip{3pt}
D_{\mu}(\mathbf{r})=-\sum_{\nu=x,y,z}\int d\mathbf{r}^\prime\frac{(\delta_{\mu\nu}-3e_\mu e_\nu)\mathbf{\textit{\textbf{f}}}_{\nu}(\mathbf{r}^\prime)}{|\mathbf{r}-\mathbf{r}^\prime|^3}
\end{equation}
In the numerical calculation, we also employ
\begin{equation}
\setlength\abovedisplayskip{3pt}
\setlength\belowdisplayskip{3pt}
\begin{split}
& D_{+}(\mathbf{r})=D_{-}^\ast (\mathbf{r})=D_{x}(\mathbf{r})+iD_{y}(\mathbf{r})\\
&                  =\frac{3}{2}\int d\mathbf{r}^\prime \frac{\sin^2\theta e^{i2\varphi}\mathbf{\textit{\textbf{f}}}_{-}(\textbf{r}^{\prime})}{|\mathbf{r}-\mathbf{r}^\prime|^3}\\
&+\frac{1}{2}\int d\mathbf{r}^{\prime} \frac{(1-3\cos^2\theta)\mathbf{\textit{\textbf{f}}}_{+}(\mathbf{r}^{\prime})}{|\mathbf{r}-\mathbf{r}^\prime|^3}+\frac{3}{2}\int d\mathbf{r}^\prime \frac{\sin{2\theta} e^{i\varphi}\mathbf{\textit{\textbf{f}}}_{z}(\mathbf{r}^\prime)}{|\mathbf{r}-\mathbf{r}^\prime|^3}
\end{split}
\end{equation}
where $\textit{\textbf{f}}_{+}=\textit{\textbf{f}}_{-}^{\ *}=\textit{\textbf{f}}_{x}+i\textit{\textbf{f}}_{y}$. The polar and azimuthal angle of the unit vector $\mathbf{e}$ are $\theta$ and $\varphi$,  respectively. For numerical convenience, we set the length unit as $a_{r}=\sqrt{\hbar/(M\omega_{0})}$, the energy unit as $E_0=\hbar\omega_{0}$ and the density unit as $n_r=N/a_r^3$.

For a set of parameters $\{U_0, \sigma\}$ of the optical plug, we can find the ground state of the dipolar spin-1 BEC by solving Eq.~(\ref{eq:gpes}) with the imaginary time propagation method~\cite{bao2012mathematical, PhysRevA.66.023613}. The phase diagram is obtained by scanning $U_0$ and $\sigma$ and the phase transition and optical plug induced vortex are then investigated. We do not expect that a tiny external magnetic field much smaller than $|\mathbf B_e|$ would change qualitatively the results.


\section{Optical plug assisted spin vortex}
\label{sec:opv}

Spin vortex in a dipolar spinor BEC has been investigated extensively in theory and experiment~\cite{sadler2006spontaneous, PhysRevLett.100.170403, PhysRevLett.97.130404, PhysRevA.81.063623, PhysRevA.82.043627}. We first briefly review a special spin vortex, the PCV~\cite{PhysRevLett.96.065302, KAWAGUCHI2012253, PhysRevA.100.033603, PhysRevResearch.3.013154, PhysRevA.94.063615}.

In general, the ground state of a $^{87}$Rb dipolar spinor BEC is a ferromagnetic SMA phase, due to $|c_2|\gg c_{dd}$ and $c_2 < 0$, where all three components share the same spatial mode. The energy of such a SMA state is
\begin{equation}\label{eq:sma}
  E^S = T + V + E_0 + E_2 +E_{dd}
\end{equation}
where $T$, $V$, and $E_0$ are the kinetic energy, the trap potential energy, and the spin-independent energy, respectively. The spin energy includes the spin exchange energy $E_2$ and the weak but long range MDDI energy $E_{dd}$. It is easy to obtain that~\cite{KAWAGUCHI2012253, PhysRevB.80.224502}
\begin{equation}\label{eq:dipE}
  E_{dd} = E_{dd}'+E_{dd}''
\end{equation}
with
\begin{eqnarray*}
  E_{dd}' &=& \frac{1}{4\pi^2}c_{dd}\int d \textbf{\textit{k}}|\hat{\textbf{\textit{k}}}\cdot\tilde{\textbf{\textit{f}}}(\textbf{\textit{k}})|^{2} \\
  E_{dd}'' &=& -\frac{2\pi}{3}{c_{dd}}\int d \textbf{\textit{r}}|{\textbf{\textit{f}}}(\textbf{\textit{r}})|^2
\end{eqnarray*}
where $\tilde {\textbf{\textit{f}}}(\mathbf k)$ is the Fourier transform of the spin density $\textbf{\textit f}(\mathbf r)$ and $\int d \textbf{\textit{k}} |\tilde{\textbf{\textit{f}}}(\textbf{\textit{k}})|^2=(2\pi)^3\int d\textbf{\textit{r}}|\textbf{\textit{f}}(\textbf{r})|^2$. One finds immediately that $E_{dd}''$ shares the same form as $E_2\equiv (c_{2}/2)\int d\textbf{r}|\textbf{\textit{f}}(\textbf{r})|^2$ so we define an equivalent spin exchange energy
$$
E_2' = E_2 + E_{dd}'' = (1+p) E_2
$$
where $p\equiv E_{dd}^{\prime\prime}/E_{2}=(4\pi c_{dd})/(3|c_{2}|)\approx 0.3797$ for a $^{87}$Rb BEC.

Similarly, for a PCV state, its energy $E^V$ is almost the same except that the energy difference from the local area around the vortex core and the dipolar energy difference from the spin structure. Clearly, for a uniform and large BEC (compared to the core size of a spin vortex), the long range dipolar energy would dominate the spin (vortex) structure because other terms are local and negligible. We thus analyze the MDDI energy difference between the SMA and the PCV state, $\Delta E =E^V-E^S \approx E_{dd}'^V-E_{dd}'^S$, where $E_{dd}'^S$ and $E_{dd}'^V$ are the intrinsic MDDI energy for the SMA and the PCV state, respectively. It is easy to find that $\Delta E < 0$ because $E_{dd}'^V = 0$, due to the circular spin density structure $\hat{\textbf{\textit{k}}}\cdot\tilde{\textbf{\textit{f}}}(\textbf{\textit{k}}) = 0$~\cite{KAWAGUCHI2012253, PhysRevLett.97.130404}, and $E_{dd}'^S > 0$. Consequently, the PCV state may become the ground state. In fact, Kawaguchi {\it et al.} found that the PCV state is indeed the ground state, if the MDDI is strong enough for a dipolar spinor condensate with the atom number larger than a threshold $N_c$~\cite{KAWAGUCHI2012253}.

Although the PCV state may be the ground state for a large enough dipolar condensate, the condition to observe such a spin texture in a finite size BEC is still quite challenging, e.g., the atom number threshold $N_c \sim 2.6\times 10^6$ for a spherical trap with $\omega_0 = (2\pi)\; 100$ Hz. Even for a disc trap with $\lambda = 20$, the threshold is still $N_c\sim 1.4 \times 10^6$. To mitigate these stringent requirements, we propose to place an optical plug at the center of the BEC. This idea is inspired by the observation that a nonzero density gradient helps the formation of the PCV structure by lowering the term $E_{dd}'$~\cite{ueda2010fundamentals}.

Let us consider a limiting case, a quasi one-dimension (1D) ring trap formed by a harmonic trap and an optical plug with a width $\sigma$ and an infinite barrier height~\cite{PhysRevA.71.033617}. Obviously, such a spin vortex is coreless since the total density is zero due to the infinite barrier height. For the FCLSV, we compare the energy difference(per atom) $\Delta E$ between the spin vortex and the SMA state. Obviously one finds
$$
\Delta E = \Delta T + \Delta E_{dd}'
$$
since other terms are the same for both states. We have defined $\Delta T = T^V-T^S$ with $T^{V,S}$ the kinetic energy for the spin vortex and the SMA state, respectively. After a straightforward calculation we find
\begin{equation}\label{eq:ring}
\Delta E = \frac{\zeta_1+\zeta_{-1}}{2\sigma^2} -\frac{c_{dd}N}{8\pi\sigma^3}I_{dd}
\end{equation}
where $\zeta_m=N_{m}/N$, with $N_{m}$ the atom number of $m$ component, and $I_{dd}$ is a constant independent of $\sigma$ defined by
$$
I_{dd} = \int_{\varphi_c}^{2\pi-\varphi_c} {d\varphi} \frac{3-2\sin^2(\varphi)}{|\sin(\varphi)|^3}
$$
with $\varphi_c$ being the cutoff (smallest) angle. The cutoff angle is in the order of $\varphi_c \sim r_0 / \sigma$ with $r_0$ the average distance between two closest atoms on the ring~\cite{PhysRevA.63.053607, dipcutoff, yukalov2018dipolar,  PhysRevA.66.023613, PhysRevA.73.041606, PhysRevLett.110.025301}. We have used $T^S=0$ (see Append.~\ref{apd:ring} for the derivation).

As shown by Eq.~(\ref{eq:ring}), the energy difference $\Delta E$ is lowered as $\sigma$ decreases, and may be negative if $\sigma$ is smaller than a characteristic width $\sigma_c ={c_{dd}NI_{dd}}/[4(\zeta_1+\zeta_{-1})]$. The ground state becomes a spin vortex. Of course, the optical plug width should be larger than the dipolar healing length $\xi_{dd}$ in order to form a vortex, $\sigma > \xi_{dd}$ with $\xi_{dd} = \hbar/\sqrt{2Mc_{dd}n}$ and $n$ the characteristic density~\cite{PhysRevLett.97.130404, PhysRevA.82.043627}. Based on the above analysis, we may draw the conclusion that the application of an optical plug may help to form a spin vortex in a dipolar spinor BEC.


\section{Truncation effect of MDDI}
\label{sec:trunc}

Accurately calculate the dipolar potential $D_{x,y,z}$ is numerically time consuming, due to the nonlocal nature of the MDDI and the multidimensional integral in the real space. A conventional method to overcome this difficulty is employing the FFT convolution theorem~\cite{PhysRevA.66.023613, shi2018variational}. However, the introduction of FFT causes quite large error because the FFT approach assumes a periodic lattice which is not a good approximation in a trapped dipolar BEC. Thus, the MDDI must be truncated to increase the numerical accuracy, particularly in the tightest trapped direction, as did in a polarized dipolar BEC~\cite{PhysRevA.74.013623}.

We restrict the dipole potential in $z$ direction with $|z-z^{\prime}|<|R_z|$ by applying a window function to the integral kernel terms with $(\delta_{\mu\nu}-3e_\mu e_\nu)/|{\mathbf r}-{\mathbf r'}|^3$~\cite{PhysRevA.82.043627, PhysRevA.63.053607}. The window function and its Fourier transform are, respectively,
\begin{equation}
W(z)=\left\{
\begin{aligned}
 & & 1 &&|z|<R_{z}\\
 & & 0 &&|z|>R_{z}
\end{aligned}
\right.\label{winr}
\end{equation}
and
\begin{equation}
\tilde{W}(k_z)=\frac{\sin(R_{z}k_z)}{\pi k_z}.
\label{wink}
\end{equation}
The kernel terms now become
\begin{equation}
	\begin{split}
		&y_{22}(\textbf{r}\!-\textbf{r}^\prime\!)= \frac{\sin^2\theta e^{i2\varphi}}{|\textbf{r}\!-\textbf{r}^\prime|^3} W(|z-z'|)\\
		&y_{21}(\textbf{r}\!-\textbf{r}^\prime\!)= \frac{\sin2\theta e^{-i\varphi}}{|\textbf{r}\!-\textbf{r}^\prime|^3} W(|z-z'|)\\
		&y_{20}(\textbf{r}\!-\textbf{r}^\prime\!)= \frac{1-3\cos^2\theta}{|\textbf{r}\!-\textbf{r}^\prime|^3} W(|z-z'|).
	\end{split}
\label{eq:dipr}
\end{equation}
After the FFT, these terms are
\begin{equation}
\begin{aligned}
\tilde{y}_{22}({\textbf{\textit k}})=&\!-\frac{4\pi}{3}\sin\alpha e^{i2\beta}\!\left[\sin\alpha\!-\!\sin\alpha \cos(R_{z}k_z)e^{\!-R_{z}k_{\perp}}\!\right.\\
&\left.+\!\cos\alpha \sin(R_{z}k_z)e^{\!-R_{z}k_{\perp}}\right ]\\
\tilde{y}_{21}({\textbf{\textit k}})\!=&\!-\frac{8\pi}{3}\sin\alpha e^{i\beta}\!\left[ \cos\alpha\!-\!\cos\alpha \cos(R_{z}k_z)e^{\!-R_{z}k_{\perp}}\!\right.\\
&\left.-\sin\alpha \sin(R_{z}k_z)e^{\!-R_{z}k_{\perp}}\right]\\
\tilde{y}_{20}({\textbf{\textit k}})\!=&\!-\!\frac{4\pi}{3}(1\!-3\cos^2\alpha)\!+\!4\pi e^{\!-R_{z}k_{\perp}}\!\left[\sin^2\alpha \cos(R_{z}k_z)\right.\\
&\left.-\!\sin\alpha \cos\alpha \sin(R_{z}k_z) \right]
\label{eq:dipk}
\end{aligned}
\end{equation}
where $k_{\perp}=\sqrt{k_{x}^2+k_{y}^2}$, $\alpha$ is the polar angle and $\beta$ the azimuthal angle of the unit vector $\hat{\textbf{\textit k}}$.

We evaluate the dipolar energy on the cubic grids of extent $[-R_x,R_x]\times [-R_y,R_y]\times [-R_z,R_z]$. The calculation box is lager than the condensate size, i.e., $R_{x,y,z}\sim1.5\times R_{x,y,z}^{c}$. The condensate size is defined as $n(R_x^c,R_y^c,R_z^c)=10^{-9}\times n_{p}$ with $n_p$ the highest density. Note that the truncation size coincides with the calculation box along $z$ axis. In order to obtain valid results, the truncation window $[-R_z,R_z]$ must be larger than the condensate size. Combining Eq.~(\ref{eq:dipk}) and Eq.~(\ref{eq:dipE}), we calculate the dipolar energy in three ways, the original, the truncated numerical method and the analytical for a given quantum state of the spin-1 condensate. The wave function has the following form $\psi_{m}(\textbf{r})=\sqrt{{n(\textbf{r})}/{3}}$ with the density
\begin{equation}
n(\textbf{r})=\sqrt{\frac{\lambda}{(2\pi)^{3}}}\exp\left[-(x^2+y^2+\lambda z^2)/2 \right].
\end{equation}
We denote the analytical MDDI energy as $E_{a}$, $E_{o}$ the original, and $E_{t}$ the truncated. The relative errors are defined as $\varepsilon_{\nu}=|(E_{\nu}-E_{a})/E_{a}|$ ($\nu=o,t)$. The results are displayed in Tab.~\ref{tab:re}. Clearly, although we may in principle reduce the relative error by simply extending the computation range with more grid points, the truncation operation is a more efficient and practical way.
\begin{table}[htbp]
\begin{center}
\caption{Relative error of the MDDI energy before and after truncation. The relative error is an order of magnitude smaller with the truncation method. The grid points are $128\times 128\times 64$ along $x,y,z$ axis, respectively.}
\scalebox{0.95}{
\begin{tabular}{l|cccccl}
\hline
\hline
$\lambda$ & 10                   &20                    & 40                   &   60            &80   \\
\hline
$R_{x,y}$ & 8                    &8                     & 11                  &   15            &22.5   \\
$R_{z}$ & 4.5                    &3.25                  & 3                  &   2.5            &1.5   \\
$\varepsilon_{o}$    & $1.4\times10^{-2}$   & $2.3\times10^{-2}$    & $1.9\times10^{-2}$   &$2.4\times10^{-2} $ &$6.9\times10^{-2} $\\
$\varepsilon_{t}$    & $4.7\times10^{-3}$   & $5.4\times10^{-3}$    & $2.4\times10^{-3}$   &$2.5\times10^{-3}$  &$9.6\times10^{-3}$ \\
\hline
\hline
\end{tabular}}
\label{tab:re}
\end{center}
\end{table}


\section{Numerical results on topological phase transition}
\label{sec:num}

\begin{figure}
\centering
\includegraphics[width=3.25in]{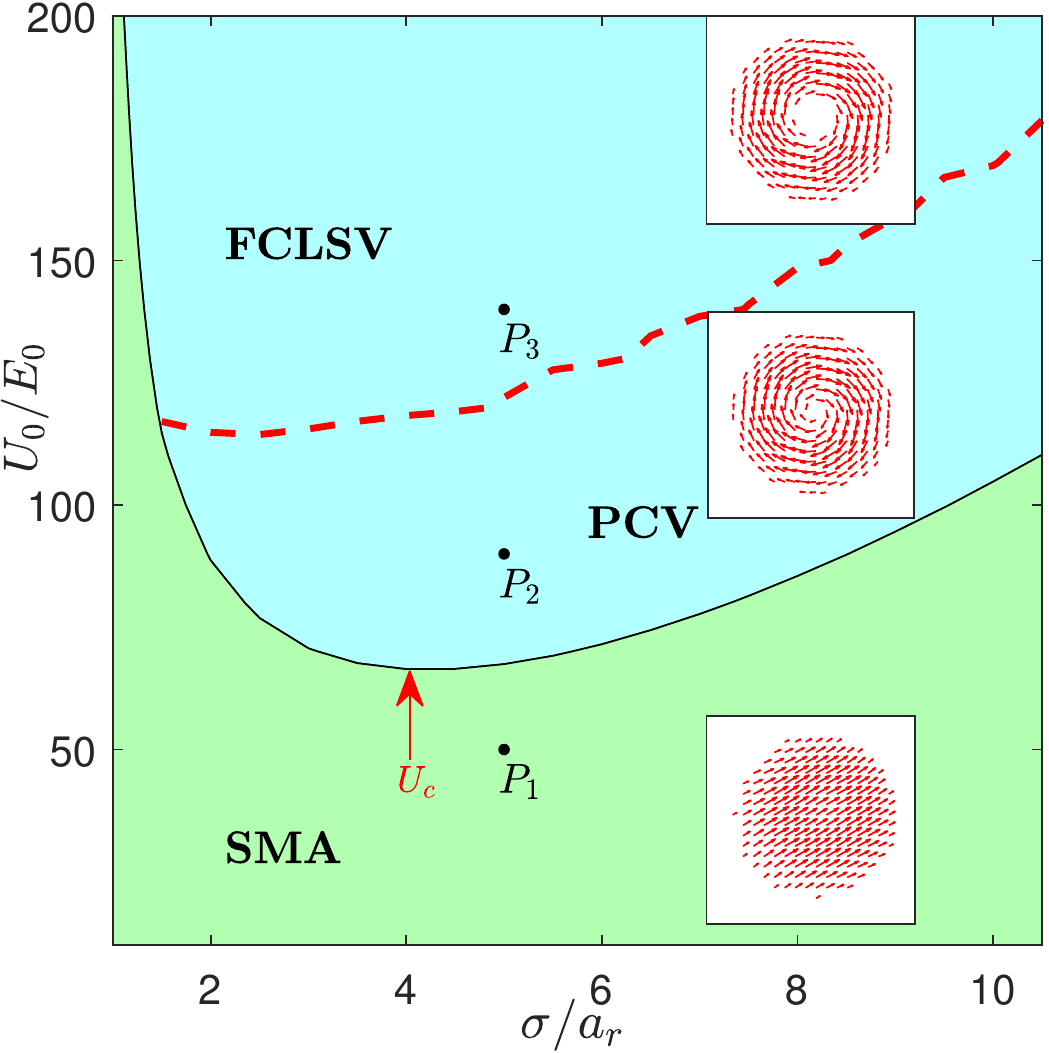}
\caption{Phase diagram of a $^{87}$Rb dipolar BEC with an optical plug. The spin vortex phase  emerges when the optical plug barrier height $U_0$ and the width $\sigma$ lie in the cyan region. Within the spin vortex phase, the FCLSV appears if the parameter $(U_0, \sigma)$ is above the dashed red line and the PCV appears below. The top inset sketches the spin density distribution in $x$-$y$ plane for the FCLSV ground state at parameter $P_{3}$. The middle inset shows that for the PCV ground state at $P_2$. The bottom inset portrays that for the SMA ground state at $P_{1}$. Other parameters are $\lambda = 20$ and $N=5\times 10^5$.}
\label{fig:pd}
\end{figure}

We determine numerically the ground state of the $^{87}$Rb dipolar spin-1 condensate by employing the conventional operator-splitting approach to evolve the three coupled GPEs Eq.~(\ref{eq:gpes}) in the imaginary time domain. The kinetic and the truncated MDDI terms are calculated with the FFT algorithm~\cite{bao2003numerical}. We set the atom number in the condensate as $N=5\times 10^5$, the trap frequency $\omega_{0}=2\pi\times 100$ Hz, the trap aspect ratio $\lambda=20$, and the calculation box $R_x=R_y=27a_r$ and $R_z=2.8a_r$, unless stated otherwise. The two optical plug parameters, the barrier height $U_0$ and the width $\sigma$, are scanned. The initial wave function is set as a parabolic shape with random coefficients.

The phase diagram in the parameter space $(U_{0},\sigma)$, as shown in Fig.~\ref{fig:pd}, summarizes the key results for the dipolar spinor $^{87}$Rb BEC. We observe that two phases are present, the SMA and the spin vortex phases, whose spin density distributions are sketched in the insets in Fig.~\ref{fig:pd}. From the figure, we find the appearance of the spin vortex phase requires the optical plug barrier height to be larger than a critical value, $U_0 > U_c$. Once the $U_0$ is large enough, the optical plug width $\sigma$ has also to be in an appropriate range, neither too big nor too small, for the spin vortex phase. This numerical result agrees qualitatively with previous analysis in Sec.~\ref{sec:opv}, where the width should lie in an intermediate range. Interestingly, we find a spin structural change within the spin vortex phase, from the PCV to the FCLSV which is coreless (nearly zero total density or numerically the total density at the center is less than $1\% $ of the peak density). The spin structural change is marked by the red dashed line in Fig.~\ref{fig:pd}. We note that this is not a phase transition because the symmetries are the same in the PCV and the FCLSV region~\cite{PhysRevA.102.013305}.

\begin{figure}
\centering
  \includegraphics[width=3.25in]{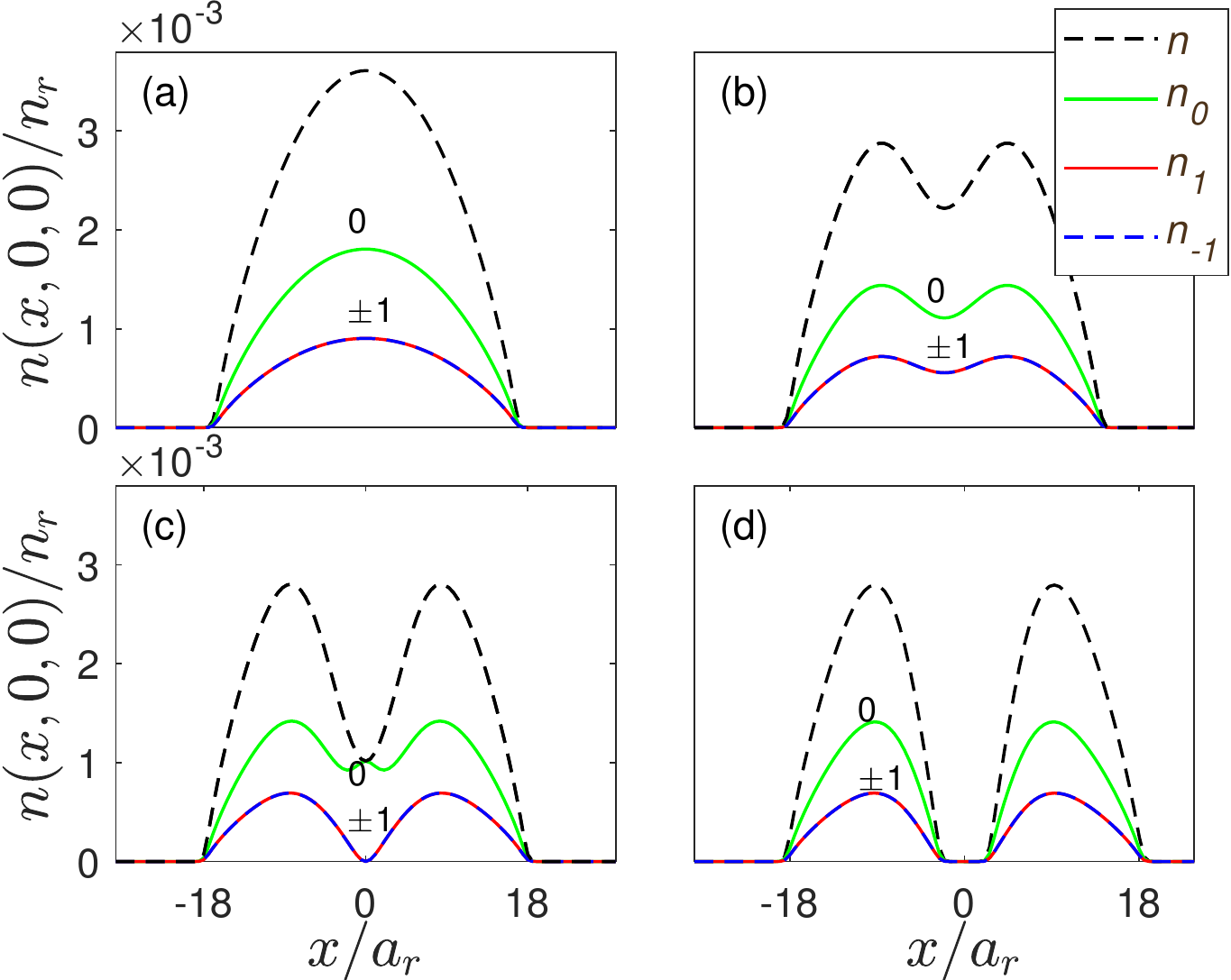}
  \caption{Density distribution of the ground state in $x$ axis ($y=0$, $z=0$) for the total density $n$ (black dashed line), three components $n_1$ (red solid line), $n_0$ (solid green line), and $n_{-1}$ (blue dashed line). The optical plug potential height is $U_0=0$ (a), $50$ (b), $90$ (c), and $140$ (d). Other parameters are $\sigma=5$, $\lambda = 20$, and $N=5\times 10^5$. (b-d) corresponds to the parameters $P_{1}$, $P_{2}$, and $P_{3}$ in Fig.~\ref{fig:pd}, respectively. (a) and (b) belong to the SMA phase where the wave function of three components share the same spatial mode. (c) and (d) belong to the PCV and the FCLSV, respectively. }
  \label{fig:density}
\end{figure}

The ground state density is significantly suppressed by the optical plug, as shown in Fig.~\ref{fig:density}. The figure shows only the density distribution along $x$-axis because the density distribution is cylindrically symmetric. As the optical plug barrier height $U_0$ increase, the total density within and around the optical plug becomes lower till nearly zero. Correspondingly, the densities of each spin component show similar trend. An interesting feature is also exhibited, i.e., the density distributions of the $|m_F = +1\rangle$ and $|m_F = -1\rangle$ components are exactly the same $n_1(\mathbf r) = n_{-1}(\mathbf r)$. The phenomenon of $f_z(\mathbf r) =n_1-n_{-1}= 0$ manifests the anisotropic property of the MDDI, i.e., spins are aligned in the easy plane which is the $x$-$y$ plane for a disc shape condensate with $\lambda \gg 1$~\cite{PhysRevLett.97.020401, PhysRevA.82.043627, PhysRevA.73.023602}. In addition, one finds that the density of the $|m_F=0\rangle$ component $n_0(\mathbf r)$ is always larger than $n_1(\mathbf r)$ and $n_{-1}(\mathbf r)$ and $n_0(\mathbf r) \approx 2n_1(\mathbf r)$ in the SMA and the FCLSV phases (not in the PCV). These features are stemmed from the requirement to lower the ferromagnetic spin exchange interaction, which reaches its lowest value if the spin is fully polarized. Combined with the property of $f_z(\mathbf r) = 0$, one immediately obtains $n_0 = 2n_1 = 2n_{-1}$ everywhere. Such a density distribution distinguishes the SMA and the FCLSV states from the PCV state where $n_0/n_{1,-1}$ is not a constant, especially around the vortex core.

\begin{figure}[!t]
\centering
  \includegraphics[width=3.25in]{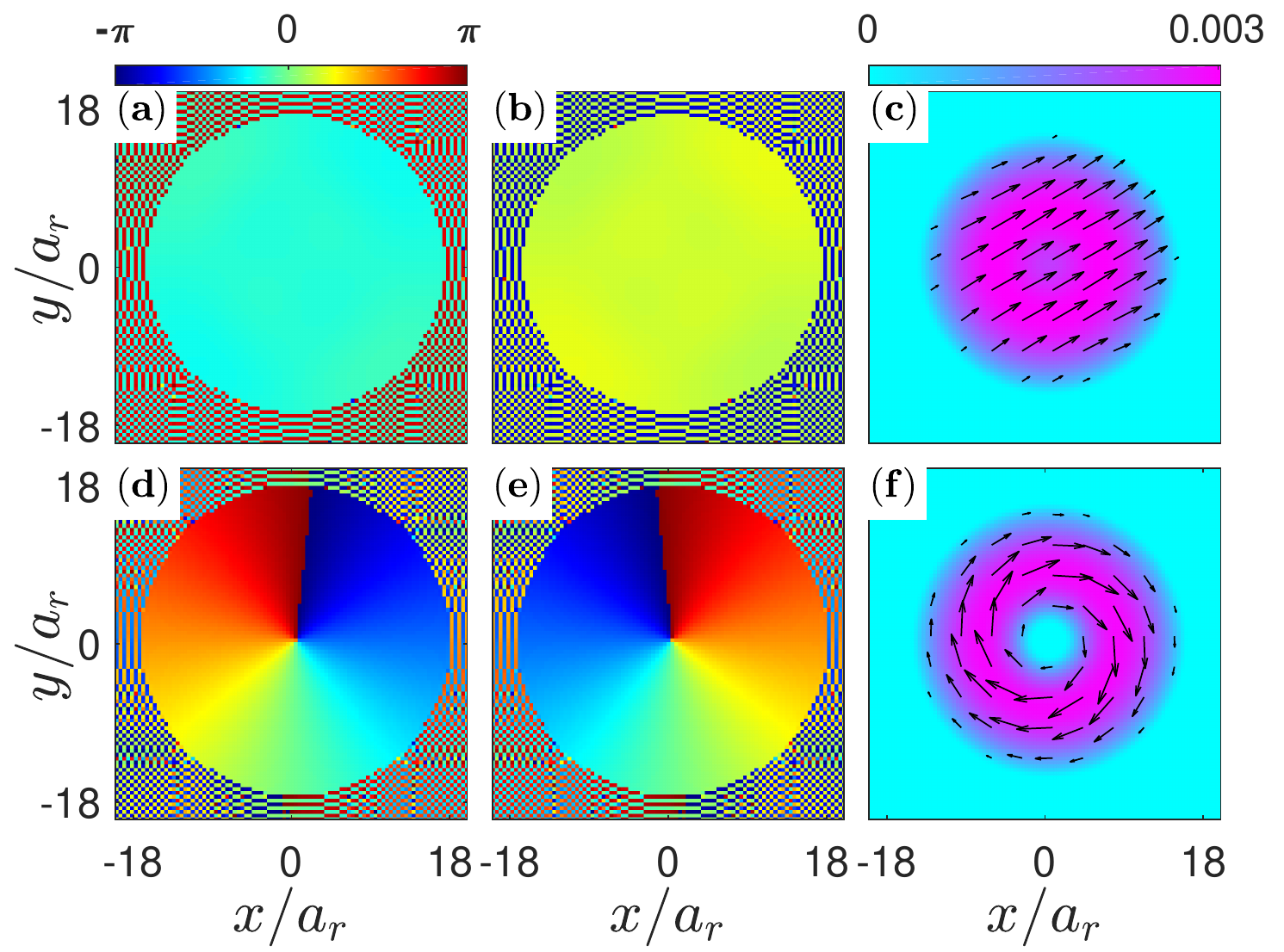}
  \caption{Spatial phase distribution of the ground state in $x$-$y$ plane ($z=0$) for $|+1\rangle$ (a) and $|-1\rangle$ component (b) with $U_0=50$ ($P_1$, SMA state  in Fig.~\ref{fig:pd} ). (d) and (e) are the same as (a) and (b) except with $U_0 = 140$ ($P_3$, FCLSV state). (c) and (f) show the corresponding spin density (arrows) at $P_1$ and $P_3$, respectively. The color denotes the total density $n$. Other parameters are the same as in Fig~\ref{fig:density}.}
  \label{fig:phase}
\end{figure}

\begin{figure}
\centering
  \includegraphics[width=3.25in]{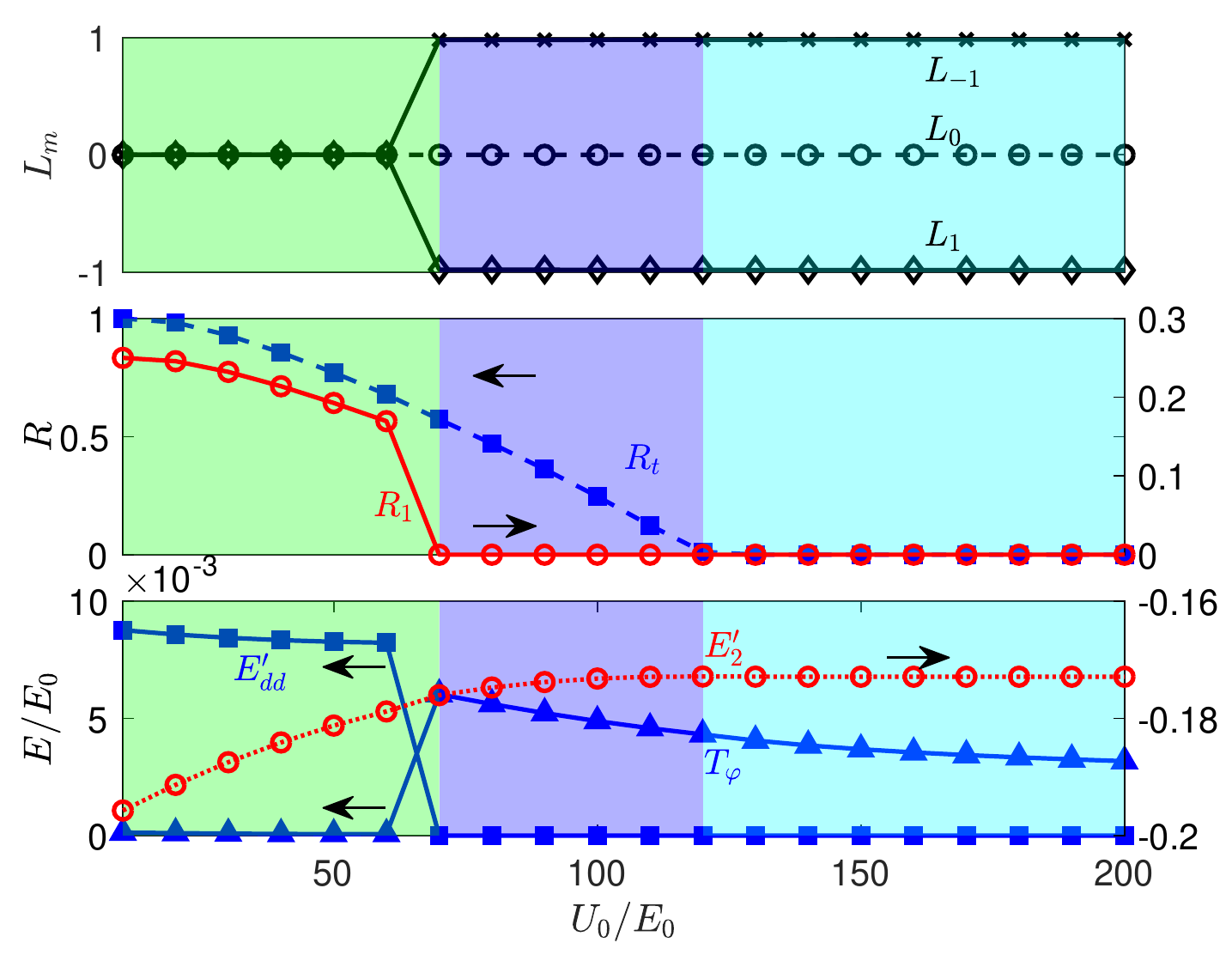}
  \caption{Phase transitions induced by optical plug with the same parameters as in Fig.~\ref{fig:density}. Phase transition from the SMA (light green) to the PCV (light blue) occurs at $U_{0}=70$, where $L_{-1}$ (solid line width crosses), $L_1$ (solid line with diamonds), $R_1$ (solid line with circles),  $E_{dd}^{\prime}$ (solid line with squares) and $T_{\varphi}$ (solid line with triangles) undergo an obvious disruption. Structure change from the PCV to the FCLSV (cyan) occurs at $U_0=120$, where $R_t$ approaches zero. $L_0$ (dashed line with circles) is always zero and $E_2^\prime$ (dotted line with circles) is continuous and smooth at both phase transitions.}
  \label{fig:Ut}
\end{figure}

Typical spatial phase and spin density distribution of a FCLSV are presented in Fig.~\ref{fig:phase}. As a comparison, we also show the constant spatial phase and spin density distribution of an SMA state. The spatial phases of $|0\rangle$ component of the FCLSV and the SMA are not shown because they are trivially constant. Clearly, the phases of an FCLSV linearly change $\mp 2\pi$ around the origin for the $|1\rangle$ and $|-1\rangle$ component, respectively, illustrating the phase distribution of a spin vortex. This phase changing is distinctive from the SMA whose phases are constant in the whole space. The spin density distribution in Fig.~\ref{fig:phase}(f) shows more direct evidence, where the local spins are aligned circularly around the center of the spin vortex. Clearly, the spin direction (along the azimuthal angle) is always perpendicular to the density gradient which has nonzero components only along radial direction and $z$-axis. Such a spin configuration guarantees $E_{dd}^{\prime}=0$ (as confirmed also in Fig.~\ref{fig:Ut}).

For a vortex, its winding number must be a nonzero integer. For the spin vortex, either the PCV or the FCLSV, we calculate the winding number of each spin component, which is proportional to the angular momentum~\cite{PhysRevLett.97.130404, PhysRevLett.96.065302},
\begin{equation}
L_{m}=\frac{\int d\textbf{r}\psi_{m}^{\ast}(\textbf{r})(-i\frac{\partial}{\partial\varphi})\psi_{m}(\textbf{r})}{\int d\textbf{r}|\psi_{m}(\textbf{r})|^2},\; (m=1, 0,-1).
\end{equation}
The $L_{m}$ is depicted in Fig.~\ref{fig:Ut} where $U_0$ increases from 0 to 200. The abrupt change of winding number from zero to $\pm 1$ is an obvious evidence of topological phase transition. For the $^{87}$Rb dipolar spinor condensate, the winding numbers are  $(-1, 0, 1)$ for the three components $|1\rangle, |0\rangle, |-1\rangle$ in the spin vortex phase where the optical plug barrier height $U_0$ is large.

The phase transition from the SMA to the spin vortex is also indicated by the density ratios for the $|\pm 1\rangle$ component $R_{1,-1}$, which is defined as $R_{1,-1} = n_{1,-1}(0,0,0)/n_p$. Since $f_z = 0$ everywhere, $n_1 = n_{-1}$ thus $R_1 = R_{-1}$ always holds. In the spin vortex phase, $R_{1,-1}=0$ indicates the densities of the $|\pm 1\rangle$ component are zero at the vortex core.

As we discussed in Sec.~\ref{sec:opv}, the spin vortex in the dipolar spinor BEC may be ascribed to the competition between the kinetic energy along the azimuthal angle
 $$T_{\varphi}=\sum_{m=-1,0,1} \int d\mathbf r \; \psi_m^*(\mathbf r)  \left(-\frac{\partial^2}{2\rho^2\partial \varphi^2}\right) \psi_m(\mathbf r)$$
and the intrinsic dipolar energy $E_{dd}^\prime$. Indeed, the intrinsic dipolar energy drops suddenly to zero but the azimuthal kinetic energy soars up from zero around the phase transition. More importantly, the sum energy $E_{dd}^{\prime} + T_{\varphi}$ also decreases abruptly once the system changes from the SMA phase to the spin vortex, provided that other spin energy terms like $E_2^\prime$ remain continuous and smooth. This confirms that the spin vortex is due to nothing but the MDDI effect.

As the optical plug potential barrier height $U_0$ further increases, the dipolar BEC changes from the PCV to the FCLSV indicated by the total density ratio $R_t$, which is denfined as $R_{t} = n(0,0,0)/n_p$. Such a gradual change in the central total density is manifested by the apprearance of the coreless vortex, i.e., $n(0,0,0) = 0$. Obviously, the FCLSV distincts itself from the PCV where a polar core exists [$n_0(0,0,0)$ is nonzero though $n_{1,-1}(0,0,0)=0$].

\begin{figure}
\centering
  \includegraphics[width=3.25in]{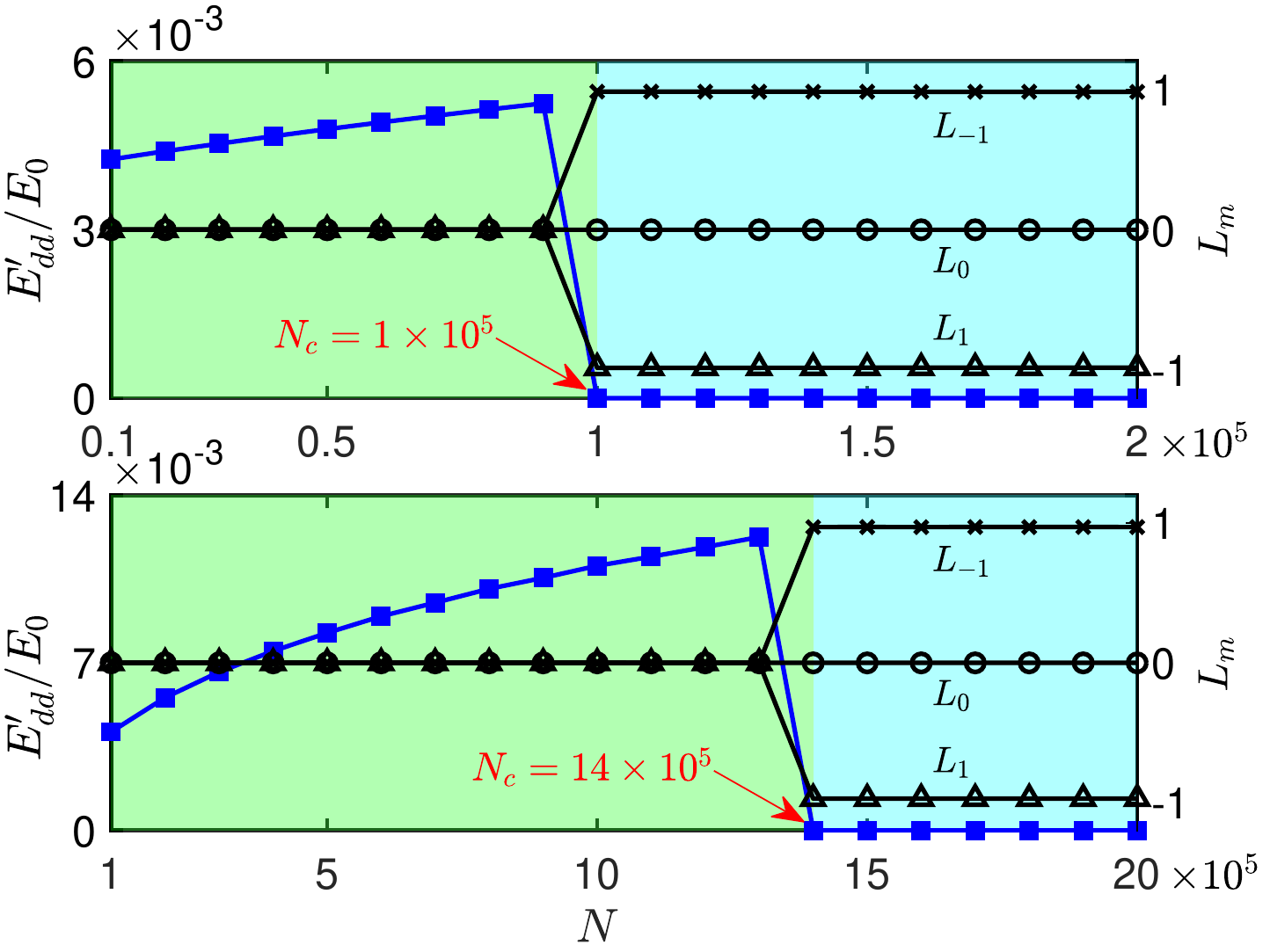}
  \caption{Same as Fig.~\ref{fig:Ut} except on the atom number $N$ for $U_0=90$ with (upper) and without (lower) an optical plug. The critical atom number $N_c$ (red arrow) with the optical plug is an order of magnitude lower than that without the optical plug.}
  \label{fig:N}
\end{figure}

More atoms in the $^{87}$Rb dipolar BEC is good to the formation of spin vortex structure~\cite{PhysRevLett.97.130404}. We investigate also the phase transition on the atom number $N$. The numerical results are shown in Fig.~\ref{fig:N}. The calculation box $R_{x,y,z}$ is adjusted appropriately with N increasing. Clearly, the spin vortex always appear if the number of atoms is larger than a critical value $N_c$, with or without an optical plug. However, the critical atom number $N_c$ with an optical plug is an order of magnitude smaller than that without the optical plug ($1.4\times10^6\rightarrow1\times10^5$). Therefore, the application of an additional optical plug mitigates greatly the experimental efforts to realize a spin vortex in the $^{87}$Rb dipolar spinor BEC.


\section{Conclusions}
\label{sec:conc}

By increasing the potential barrier height and adjusting appropriately the width of an additionally applied optical plug, a $^{87}$Rb dipolar spinor condensate transits from a spin-uniform SMA phase to a spin vortex one, due to the competition between the MDDI and the azimuthal kinetic energy. With the aid of the optical plug, it is possible to generate the PCV and the FCLSV states of the $^{87}$Rb dipolar condensate in a controllable way under more relaxed experimental conditions, e.g., an order of magnitude less atom number and smaller aspect ratio. Our results provides a practical way to realize the spin vortex state and to explore the topological quantum phase transition and dynamical phase transitions in $^{87}$Rb condensates.

Recent experiments in ferromagnetic spin-1 $^7$Li BEC demonstrate a very large spin-dependent exchange interaction strength ($|c_2|\sim c_0/2$)~\cite{PhysRevResearch.2.033471, PhysRevLett.127.113001, PhysRevLett.127.043401}. For such strong ferromagnetic interaction, it would be more challenging to generate a spin vortex state. However, as implied by Eq.~(\ref{eq:ring}), a FCLSV state is still possible with the assistance of an optical plug. It is worthy to explore this novel regime in the future.


\begin{acknowledgments}
This work is supported by the National Natural Science Foundation of China under Grant Nos. U1930201, 12135018 and 91836101.
\end{acknowledgments}

\appendix
\section{Derivation of Eq.~(\ref{eq:ring})}
\label{apd:ring}

In the quasi one-dimension (1D) ring trap, we assume that the ring of BEC lies in 2D plane with a linear density $n={N}/({2\pi \sigma})$, where $\sigma$ is the ring's radius. For the $^{87}$Rb dipolar BEC, the 1D wave function is expressed as $\psi_{m}(\varphi)=(1/\sqrt{2\pi})\exp(i\theta_m)$  ($m=1,0,-1$), where the phase $\theta_m=k_{m}\varphi+\alpha_m$ satisfies~\cite{PhysRevA.66.011601, doi:10.1143/JPSJ.70.1604}
\begin{equation}
	\setlength\abovedisplayskip{3pt}
\theta_{1}+\theta_{-1}-2\theta_{0}=0.
	\setlength\belowdisplayskip{3pt}
\end{equation}
For the spin vortex we consider, $k_m=-m$, and for the SMA phase $k_m=0$~\cite{PhysRevLett.97.020401}.

We calculate the energy difference $\Delta E$ between two local-energy-minimum states, the SMA and the spin vortex state, denoted by the superscript $S$ and $V$, respectively. Due to the homogeneous $1$D density, the potential energy difference $\Delta V\equiv V^{V}-V^{S}=0$ and the spin-independent (density) interaction energy difference $\Delta E_{0}=0$. The kinetic energy is
\begin{equation}
 	 		\setlength\abovedisplayskip{3pt}
 T=\sum_{m}\int \sigma d\varphi \psi_{m}^{\ast}\left(-\frac{1}{2\sigma^2} \frac{\partial^2}{\partial\varphi^2}\right) \psi_{m} =\frac{1}{2\sigma^2}\sum_{m}\zeta_{m}k_{m}^{2}
 \end{equation}
where $N_{m}$ is the atom number of the $m$ component and $\zeta_m=N_m/N$. Clearly, $T^S = 0$ thus the kinetic energy difference $\triangle T=T^{V}-T^{S}=T^{V}=(\zeta_1+\zeta_{-1})/({2\sigma^2})$.

To calculate the MDDI energy difference, we cutoff the dipolar interaction at the average distance of two neighboring atoms $r_c=2\pi \sigma/N$~\cite{PhysRevA.63.053607}, which yields the cutoff of the azimuthal angle $\varphi_c=2\pi/N$ on the ring. The MDDI energy difference is
\begin{equation}
\setlength\abovedisplayskip{3pt}
\begin{aligned}
&\triangle E_{dd}=E_{dd}^{V}-E_{dd}^{S}\\
&=-\frac{c_{dd}N}{16\pi^2\sigma^3}\!\int_{\varphi_i}^{\varphi_f} \!d\varphi_{-}\int_{0}^{2\pi}\! d\varphi_{+}\!\; \frac{3\!-\!2\sin^2(\varphi_{-})\!+3\cos(2\varphi_{+})}{|\sin(\varphi_{-})|^3}\\
&=-\frac{c_{dd}N}{8\pi\sigma^3}\int_{\varphi_i}^{\varphi_f} {d\varphi_{-}}\frac{3-2\sin^2(\varphi_{-})}{|\sin(\varphi_{-})|^3}\\
&=-\frac{c_{dd}N}{8\pi\sigma^3}I_{dd}
\end{aligned}
\end{equation}
where $\varphi_i=\varphi_c$, $\varphi_f=2\pi-\varphi_c$, $\varphi_{\pm}=(\varphi\pm\varphi^{\prime})/2$ with $\varphi$ and $\varphi^{\prime}$ representing the azimuthal angle of two atoms located at $(\sigma,\varphi)$ and $(\sigma,\varphi^{\prime})$ respectively. Obviously, $I_{dd}$ is only a positive number determined by the atom number and the truncation $\varphi_c$. As a result, the total energy difference becomes
\begin{equation}
\setlength\abovedisplayskip{3pt}
\begin{aligned}
\triangle E&=\triangle T+\triangle E_{dd}\\
&= \frac{\zeta_1+\zeta_{-1}}{2\sigma^2} -\frac{c_{dd}N}{8\pi\sigma^3}I_{dd}.
\end{aligned}
\end{equation}


\end{document}